# Dummy-Atom-Based Thermodynamic Integration Accurately Predicts Disulfide Redox Free Energies in Proteins

Daniel Mejia-Rodriguez,[#a] Suman Samantray,[#a] Margaret S. Cheung,[b,c] and Hoshin Kim[*a]

[a]Physical Sciences Division, Physical and Computational Sciences Directorate, Pacific Northwest National Laboratory, Richland, Washington, USA

[b]Environmental Molecular Sciences Laboratory, Richland, Washington, USA

[c]University of Washington, Seattle, Washington, USA

[#]shares co-first authorship of this article.

[*]Corresponding Author: hoshin.kim@pnnl.gov

Supplementary Information available: [Detailed bonded and non-bonded parameters transforming from state A to B, details of experimental and calculated redox potentials and free energies of tested proteins, and changes in S-S distance during the TI with a different bonded parameter]

**Abstract**

**A thermodynamic integration (TI) protocol incorporating dummy atoms is introduced to calculate free energy differences (ΔG) for disulfide bond formation in proteins. This method successfully reproduces experimental redox potentials for multiple proteins, providing improved insights into the redox regulation of various proteins.**

Redox regulations are a central process in biology, governing a wide range of protein functions, including energy production, respiration-photosynthesis, or cellular signalling.[1, 2] Among amino acids, cysteine residues frequently participate in redox reactions via post-translational modifications (PTMs), with disulfide bond formation being one of the most common.[3] These bonds are formed through the covalent linkage of two cysteine residues and the oxidation of the sulfhydryl (SH) groups. Because of their critical biological role, the accurate evaluation of redox potentials and free energy changes associated with disulfide bond formation and breakage has become a major focus.[4] While experimental methods can determine redox potentials or free energy changes regarding disulfide bond formation with high accuracy, their applicability is often limited by partner dependence, experimental conditions, the complexity of mutagenesis, and the resource demands.[5] Consequently, computational approaches have become important for reliable and efficient predictions. Recently, force-field–based molecular dynamics (MD) simulations combined with the Crooks Gaussian Intersection (CGI) method have been developed to predict disulfide redox potentials of various proteins and their mutants with improved accuracy compared to the previous approaches.[5, 6] In this approach, dummy atoms were introduced to model sulfur protonation while converting disulfide-linked cysteines into two separate residues.

In this study, inspired by the previously reported CGI method,[5] we applied a dummy atom approach within a thermodynamic integration (TI) framework. Using the same test cases as the earlier work, together with a larger protein for which experimental redox potential data are available, we observed a substantial improvement in accuracy, although some limitations remain that should be addressed for further refinement. To facilitate broader use, we also automated the process of introducing dummy atoms after MD simulations, running, analyzing, and post-processing the TI calculations, and implemented it in our newly developed Python package, PTM-Psi.[7, 8]

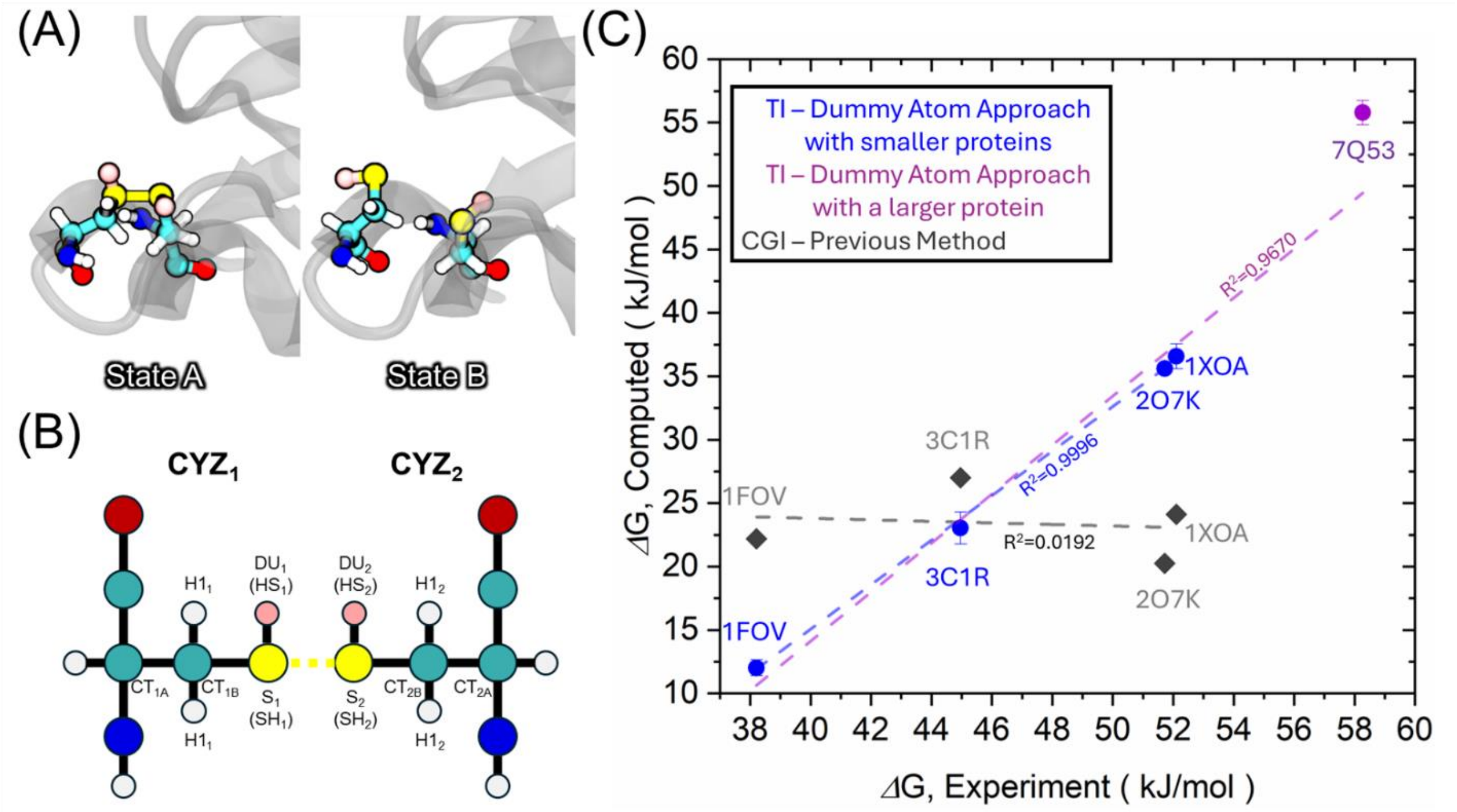

**Figure 1. (A)** Representative snapshots of before (state A) and after (state B) the TI. CYZ residues are highlighted using a CPK model. **(B)** Scheme of the customized disulfide-bonding cysteine residue (CYZ). Colors denote hydrogen (white), carbon (cyan), oxygen (red), nitrogen (blue), sulfur (yellow), and dummy atom (pink) in the initial state. Atom types undergoing transformation during TI are labeled, with bracketed types indicating the final state. **(C)** Comparison of experimental and computed binding free energies. Gray, blue, and purple data points with regression lines ($R^2$ values shown) correspond to the CGI approach,5 dummy-atom-based TI with smaller proteins, and dummy-atom-based TI with a larger protein, respectively.

The redox potential for a given electrochemical reaction can be evaluated by first computing the associated free energy change, because the maximum amount of work producible by such a reaction is equal to the product of the cell potential and the total charge transferred during the reaction. In this work, we employed MD simulations using the GROMACS software package[9] to obtain the desired free energy change upon disulfide bond reduction by TI along an alchemical pathway. The proposed protocol uses the dual topology approach,[10] whereby the parameters of the cystine dimer (the oxidized, disulfide-bonded form of cysteine) in state A are gradually transformed into those of a regular cysteine (CYS) in state B (**Figure 1A**). To accomplish such a transformation, we defined a new residue type called CYZ in the Amber ff99SB[11] and ff14SB force fields[12] implemented in GROMACS. The CYZ residue type is analogous to the standard deprotonated cysteine involved in disulfide bonds, so-called CYX, in the sense that both are used to describe the cystine dimer. The new CYZ residue also inherits atomic types and charges for all atoms present in CYX but adds two dummy hydrogen atoms with zero charge (**Figure 1B**). To facilitate the user experience, we added a new protocol to the PTM-Psi Python package[7] that, given a PDB file (1) identifies cystine dimers by S-S distance, (2) inserts the dummy hydrogen atoms using the Natural Extension Reference Frame (NeRF) algorithm,[13] (3) determines the number of clashes of the new dummy atoms with all others within a sphere with 10 Å radius, (4) changes the position of the new dummy atoms to minimize the number of clashes if detected, (5) changes the residue name from CYS/CYX to CYZ, and (6) writes a new PDB file with the updated information. It is important to note that, since the dummy atoms do not influence the dynamics of the system, this new protocol can be used on a crystal/inferred three-dimensional structure to start the phase space sampling from scratch, or on a representative conformation from a previous MD trajectory. The clash search algorithm circumvents the formal $N_{atom}^2$ computational effort by first comparing the distances between alpha carbon atoms, resulting in an algorithm with roughly $N_{res}^2$ scaling. The PTM-Psi package was also extended to facilitate the setup of the TI free energy calculations involving the reduction of disulfide bonds. This extension reads an existing GROMACS topology file that already includes the CYZ residue type and adds parameters of state B for free cysteine, including updated charges, atom types, and bonded force constants. The script also adds the missing 1-2, 1-3, and 1-4 short-range interactions, as well as some missing long-range van der Waals (vdW) interactions, to state B. These interactions are missing in state A due to the presence of the S-S bond. The detailed differences in bonded parameters, 1-2, 1-3, 1-4 interactions, and long-range interactions between State A and B are provided in **Tables S1** and **S2** of the Supporting Information. To test our approach in comparison with the previous CGI method,[5] we computed the free energy change upon disulfide bond reduction of five proteins for which either their experimental or computed redox potentials have been reported. Particularly, we selected the four proteins from the previous work that showed the least agreement to experimental data, including glutaredoxin (Grx) 3 from *Escherichia coli* (PDB ID: 1FOV, 82 residues), [14] Grx1 from *Escherichia coli* (PDB ID: 3C1R, 118 residues),[15] thioredoxin (Trx) from *Staphylococcus aureus* (PDB ID: 2O7K, 107 residues),[16] Trx from Escherichia coli (PDB ID: 1XOA, 108 residues).[17] Also, we selected the $A_2B_2$ tetramer form of the glyceraldehyde-3-phosphate dehydrogenase (GAPDH) from *Spinacia oleracea* (PDB ID: 7Q53, 1352 residues).[18] Hereafter, we refer to the proteins by their PDB IDs. All MD simulations start from the oxidized form (disulfide bond present CYZ) and were prepared with the default settings of the PTM-Psi package (see Refs. 7 and 8 for details). In short, all titratable residues were protonated according to a pH value of 7.0, solvated with TIP3P water in a truncated dodecahedron box with 1 nm padding, and neutralized with as many Na+ and Cl- ions as needed to achieve 0.154 M salt concentration. A stepwise energy minimization and equilibration procedure, with decreasing heavy-atom restraints, was then used to ensure smooth equilibration to 300 K and 1 bar. Bonds to hydrogen atoms are constrained using the LINCS algorithm,[19] and the time step is fixed at 2 fs. After 100 ns sampling in the NPT ensemble, a representative structure is selected, and the TI protocol is started using 30 λ values for the coupling parameter, with smaller λ-steps towards State B (free cysteines). During the transformation from State A to State B along λ, the first 13 steps were used to gradually change electrostatic interactions (Q) (λ = 0.00, 0.05, 0.10, 0.20 ... 0.90, 0.95, 1.00). Subsequently, beginning at step 14, vdW and bonded interactions varied in a similar manner over the next 13 steps. To potentially decrease the appearance of rather large fluctuations due to the softening (and eventually breaking) of the S-S bond, additional λ of 4 more steps (0.99, 0.993, 0.996, 0.999, and 1.00, to the alchemical transformation pathway. There is a total of 30 steps for the stepwise transformation of both nonbonded and bonded terms. We perform short simulations (1 ns with 1 fs time step) at each λ-step to extract the instantaneous $\partial_\lambda U$, average them, and obtain the desired free energy change as

$$\Delta G = \int_0^1 \left\langle \frac{\partial U}{\partial \lambda} \right\rangle d\lambda$$

The Bennett's Acceptance Ratio (BAR) method[20] is used to reduce the integration bias because the reduction of the disulfide bond leads to asymmetric end-point fluctuations. Results obtained from experiments or calculations were converted between free energy and redox potential using the Nernst equation:

$$-\Delta G = EnF$$

where $\Delta G$ and E are the free energy differences and redox potential between the two states in redox reactions, respectively, n is the number of transferred electrons, and F is the Faraday constant. Furthermore, the experimental reduction potential of protein 7Q53, determined at pH 7.9, was adjusted to pH 7.0 using the Nernst equation for the standard hydrogen electrode, which in millivolts reads

$$\Delta E = -59.1 \Delta pH$$

First, we performed the MD simulations and applied our TI-based approach to four small proteins (molecular weight less than 15 kDa) for which both their experimental and computed redox potentials have been reported. These include 1XOA, 2O7K, 3C1R, and 1FOV. These enzymes are known to be ubiquitous proteins that reduce disulfide bonds for their redox regulation and have been used as benchmarks for the previously reported CGI approach.[5] In this previous study, multiple proteins and their mutants were tested; among them, we chose four wild-type proteins that displayed the least agreement with the experimental trend, showing a notably low regression slope. As illustrated in **Figure 1C** and **Table S3**, no correlations were observed between experimental $\Delta G$ values or redox potentials[16, 17, 21] and their computed values calculated with the CGI approach ($R^2$ = 0.0192).5 In contrast, our dummy-atom-based TI method produced $\Delta G$ values that were linearly correlated with experimental data, yielding an $R^2$ of 0.9996. Despite this strong correlation, discrepancies remain between the absolute free energy change computed with our approach and the corresponding experimental data (it is important to note that **Figure 1C** uses different ranges for the X and Y axes). These differences likely stem from certain limitations of the current approach, which we acknowledge but cannot yet address due to technical constraints. First, only vdW interactions were transformed from short- (e.g., 1-3 or 1-4 interactions) to long-range contributions during the TI; and short-range Q were left unchanged. For example, $CT_{1B}$ and $CT_{2B}$ were treated as 1-4 interactions for both Q and vdW, but only vdW terms were switched to long-range contributions during the TI. This stems from the limitations in GROMACS, which does not allow dual-topology alchemical transformations involving Q to be transformed from short- to long-range interactions. It is plausible that this factor accounts for the relatively large discrepancies between experimental and computed absolute free energy changes in the small protein systems. Previous studies applied a linear fit with a residual error to shift the redox potential to the experimental values;[5, 6] however, this correction may not be rational, as it could be system-dependent. Interestingly, the discrepancies diminished when our method was applied to larger proteins, showcasing the system-dependent nature mentioned above. For example, in the case of GAPDH from spinach, 7Q53 (molecular weight = 147.7 kDa), where redox regulation plays a critical role in photosynthesis, the computed ΔG agreed with the experimental value[22, 23] within a ~3 kJ/mol deviation (**Figure 1C** and **Table S3**). Also, the correlation between experiment and simulation remained strong ($R^2$ = 0.9670) upon inclusion of the additional data point (**Figure 1C**). This suggests that, in larger proteins, the effect of local short- vs. long-range interactions of cysteine residues during disulfide bond reduction could become negligible, as these residues are embedded within a more complex network of nonbonded interactions. Because force field-based MD simulations are not capable of describing covalent bond dissociation, it is essential in our approach to define the equilibrium bond distance (b0) value for the distance between two sulfur atoms (S-S) a priori. The choice of this $b_0$ can influence the conformation between the two cysteine residues, which could consequently impact the computed ΔG values. In our approach, the S-S distance consistently converged to a specific value in the final λ windows, regardless of the initial b0 value. For example, the simulations of 1XOA showed that, independent of the chosen b0, the distance converged to approximately 3.6 Å, as the system approached State B. In particular, this converged value was found to be in line with the S-S distance (3.67 Å) measured in the crystal structure of its reduced form (PDB ID: 1XOB), which contains two reduced cysteines (**Figure S1**). This close agreement strongly supports the validity of our approach in capturing the structural features associated with disulfide bond reduction.

Our proposed method offers a practical and accessible approach for estimating redox potentials and free energy differences for disulfide bond formations in proteins. Because it is already implemented in the PTM-Psi package, it can be readily applied by other users. The approach involves attaching dummy atoms prior to the TI calculations, meaning that new MD simulations do not need to be performed to introduce dummy atoms. While its accuracy may not match that of the level of QM-based approaches, our method provides results more efficiently, making it particularly attractive for large-scale or exploratory studies. Nevertheless, two challenges remain. First, as aforementioned, the current implementation does not fully account for all non-bonded interactions during alchemical transformations. Second, the choice of the equilibrium bond distances requires careful fine-tuning. Addressing these limitations is expected to further improve the accuracy and robustness of the method.

In summary, our new dummy-atom-based TI approach, inspired by the earlier CGI method,[5] shows better agreement with experimental data and offers a practical alchemical framework for estimating redox potentials and free energy differences between proteins with reduced vs. disulfide-bonded cysteine residues. Moreover, since this method has already been automated and integrated into the PTM-Psi package,[7] it offers the advantage of being readily applicable without additional implementation effort. Ultimately, this approach has the potential to serve as an effective tool for probing how PTMs and redox changes influence protein structure and function, thereby advancing our understanding of redox regulation in biology.

This research was funded by the Generative AI for Science, Energy, and Security Science & Technology Investment under the Laboratory Directed Research and Development Program at Pacific Northwest National Laboratory (PNNL). This work was also supported by the Center for AI and Cloud Computing at PNNL. This work was partially supported by the NW-BRaVE for Biopreparedness project funded by the U. S. Department of Energy (DOE), Office of Science, Biological and Environmental Research program, under FWP 81832. A portion of this research was performed on a project award (61054) from the Environmental Molecular Sciences Laboratory, a DOE Office of Science User Facility sponsored by the Biological and Environmental Research program. PNNL is a multi-program national laboratory operated by Battelle for the DOE under Contract No. DE-AC05-76RL01830.

## Conflicts of interest

There are no conflicts to declare.

## Data availability

The code for the PTM-Psi version used during this study can be found at https://github.com/pnnl/PTMPSI under the protected branch TI_disulfides.

Initial configurations, subsampled trajectories, log files, and portable binary run input (tpr) files are deposited at NOMAD (https://doi.org/10.17172/NOMAD/2025.09.16-1). Raw MD trajectories for all systems can be provided upon request.

## References

1. H. Sies, R. J. Mailloux and U. Jakob, Nature Reviews Molecular Cell Biology, 2024, 25, 701–719.
2. H. Kim, S. Feng, P. Bohutskyi, X. Li, D. Mejia-Rodriguez, T. Zhang, W.-J. Qian and M. S. Cheung, arXiv [physics.bio-ph], 2025.
3. N. M. Giles, A. B. Watts, G. I. Giles, F. H. Fry, J. A. Littlechild and C. Jacob, Chem Biol, 2003, 10, 677–693.
4. M. Huber-Wunderlich and R. Glockshuber, Folding and Design, 1998, 3, 161–171.
5. W. Li, I. B. Baldus and F. Gräter, The Journal of Physical Chemistry B, 2015, 119, 5386–5391.
6. L. Lin, H. Zou, W. Li, L. Y. Xu, E. M. Li and G. Dong, Front Chem, 2021, 9, 797036.
7. D. Mejia-Rodriguez, H. Kim, N. Sadler, X. Li, P. Bohutskyi, M. Valiev, W. J. Qian and M. S. Cheung, Protein Sci, 2023, 32, e4822.
8. S. Samantray, M. Lockwood, A. Andersen, H. Kim, P. Rigor, M. S. Cheung and D. Mejia-Rodriguez, arXiv [physics.bio-ph], 2025.
9. M. J. Abraham, T. Murtola, R. Schulz, S. Páll, J. C. Smith, B. Hess and E. Lindahl, SoftwareX, 2015, 1-2, 19–25.

10. T. P. Straatsma and J. A. McCammon, The Journal of Chemical Physics, 1991, 95, 1175–1188.
11. V. Hornak, R. Abel, A. Okur, B. Strockbine, A. Roitberg and C. Simmerling, Proteins, 2006, 65, 712–725.
12. J. A. Maier, C. Martinez, K. Kasavajhala, L. Wickstrom, K. E. Hauser and C. Simmerling, J Chem Theory Comput, 2015, 11, 3696–3713.
13. J. Parsons, J. B. Holmes, J. M. Rojas, J. Tsai and C. E. M. Strauss, J Comput Chem, 2005, 26, 1063–1068.
14. K. Nordstrand, A. Sandström, F. Åslund, A. Holmgren, G. Otting and K. D. Berndt, J Mol Biol, 2000, 303, 423–432.
15. J. Yu, N.-N. Zhang, P.-D. Yin, P.-X. Cui and C.-Z. Zhou, Proteins: Structure, Function, and Bioinformatics, 2008, 72, 1077–1083.
16. G. Roos, A. Garcia-Pino, K. Van belle, E. Brosens, K. Wahni, G. Vandenbussche, L. Wyns, R. Loris and J. Messens, J Mol Biol, 2007, 368, 800–811.
17. M.-F. Jeng, A. P. Campbell, T. Begley, A. Holmgren, D. A. Case, P. E. Wright and H. J. Dyson, Structure, 1994, 2, 853–868.
18. R. Marotta, A. Del Giudice, L. Gurrieri, S. Fanti, P. Swuec, L. Galantini, G. Falini, P. Trost, S. Fermani and F. Sparla, Acta Crystallographica Section D, 2022, 78, 1399–1411.
19. B. Hess, H. Bekker, H. J. C. Berendsen and J. G. E. M. Fraaije, J Comput Chem, 1997, 18, 1463–1472.
20. C. H. Bennett, J Comput Phys, 1976, 22, 245–268.
21. F. Åslund, K. D. Berndt and A. Holmgren, J Biol Chem, 1997, 272, 30780–30786.
22. F. Sparla, P. Pupillo and P. Trost, J Biol Chem, 2002, 277, 44946–44952.
23. L. Gurrieri, F. Sparla, M. Zaffagnini and P. Trost, Seminars in Cell  Developmental Biology, 2024, 155, 48–58

# Supporting Information

# Dummy-Atom Based Thermodynamic Integration Accurately Predicts Disulfide Redox Free Energies in Proteins

Daniel Mejia-Rodriguez,[1,#] Suman Samantray,[1,#] Margaret S. Cheung,[2] and Hoshin Kim[1,*]

[1]Physical Sciences Division, Physical and Computational Sciences Directorate, Pacific Northwest National Laboratory, Richland, Washington, USA

[2]Environmental Molecular Sciences Laboratory, Richland, Washington, USA

[3]University of Washington, Seattle, Washington, USA

[#]shares co-first authorship of this article.

[*]Corresponding Author: hoshin.kim@pnnl.gov

**Table S1.** Bonded Parameters transforming from the state A (two CYZs with dummy atoms) to the state B (two separate CYSs). Unit of parameters are following: $b_0$ (nm), $k_b$ (kJ $mol^{-1}$ $nm^{-2}$), $\theta_0$ (deg.), $k_\theta$ (kJ $mol^{-1}$ $rad^{-2}$), phase (deg.), $K_d$ (kJ $mol^{-1}$), and multiplicity (pn, unitless).

<table>
<tr><th>Bonds</th><th>$b_0$ | $k_b$<br>(State A → B )</th><th>Torsions</th><th>Phase | $K_d$ | pn<br>(State A → B)</th></tr>
<tr><td>S-DU</td><td>0.133 | 229283.2</td><td>CT-S-S-CT</td><td>0.0 | 14.64400 | 2</td></tr>
<tr><td>SH-HS</td><td>0.133 | 229283.2</td><td>CT-SH SH-CT</td><td>0.0 | 00.00000 | 2</td></tr>
<tr><td>S-S</td><td>0.204 | 138908.8</td><td>CT-S-S-CT</td><td>0.0 | 2.51040 | 3</td></tr>
<tr><td>S S</td><td>0.450 | 000000.0</td><td>CT-SH SH-CT</td><td>0.0 | 0.00000 | 3</td></tr>
<tr><td>CT-S</td><td>0.181 | 189953.6</td><td>DU-S-S-DU</td><td>0.0 | 0.00000 | 2</td></tr>
<tr><td>CT-SH</td><td>0.181 | 198321.6</td><td>DU-SH SH-DU</td><td>0.0 | 0.00000 | 2</td></tr>
<tr><td>Angles</td><td>$\theta_0$ | $k_\theta$<br>(State A → B)</td><td>DU-S-S-DU<br>DU-SH SH-DU</td><td>0.0 | 0.00000 | 3<br>0.0 | 0.00000 | 3</td></tr>
<tr><td>CT-S-DU</td><td>96.0 | 000.000</td><td>DU-S-S-CT</td><td>0.0 | 0.00000 | 2</td></tr>
<tr><td>CT-SH-HS</td><td>96.0 | 359.824</td><td>DU-SH SH-CT</td><td>0.0 | 0.00000 | 2</td></tr>
<tr><td>CT-S-S</td><td>103.7 | 569.024</td><td>DU-S-S-CT</td><td>0.0 | 0.00000 | 3</td></tr>
<tr><td>CT-S S</td><td>103.7 | 000.000</td><td>DU-SH SH-CT</td><td>0.0 | 0.00000 | 3</td></tr>
<tr><td>DU-S-S</td><td>103.7 | 282.280</td><td>H1-CT-S-S</td><td>0.0 | 1.39467 | 3</td></tr>
<tr><td>HS-SH SH</td><td>103.7 | 000.000</td><td>H1-CT-SH SH</td><td>0.0 | 0.00000 | 3</td></tr>
<tr><td>H1-CT-S</td><td>109.5 | 418.400</td><td>CT-CT-S-S</td><td>0.0 | 1.39467 | 3</td></tr>
<tr><td>H1-CT-SH</td><td>109.5 | 418.400</td><td>CT-CT-SH SH</td><td>0.0 | 0.00000 | 3</td></tr>
<tr><td>CT-CT-S</td><td>114.7 | 418.400</td><td>H1-CT-S-DU</td><td>0.0 | 0.00000 | 3</td></tr>
<tr><td>CT-CT-SH</td><td>108.6 | 418.400</td><td>H1-CT-SH-HS</td><td>0.0 | 1.04600 | 3</td></tr>
</table>

**Table S2.** A list of non-bonded interactions that undergo transformations during the TI. OFF and ON indicates given non-bonded interactions are ignored or considered. 1-4 means that the non-bonded interactions are considered as 1-4 interactions.

| Atom 1 | Atom 2 | State A (disulfide) | State B (2CYSs) |
|---|---|---|---|
| 1-2 interactions | | | |
| $S_1(SH_1)$ | $S_2(SH_2)$ | OFF | ON |
| 1-3 interactions | | | |
| $S_1(SH_1)$ | $CT_{2B}$ | OFF | ON |
| $S_2(SH_2)$ | $CT_{1B}$ | OFF | ON |
| $DU_1(HS_1)$ | $S_2(HS_2)$ | OFF | ON |
| $DU_2(HS_2)$ | $S_1(HS_1)$ | OFF | ON |
| 1-4 interactions | | | |
| $DU_1(HS_1)$ | $CT_{1A}$ | OFF | 1-4 |
| $DU_2(HS_2)$ | $CT_{2A}$ | OFF | 1-4 |
| $DU_1(HS_1)$ | $H1_1$ | OFF | 1-4 |
| $DU_2(HS_2)$ | $H1_2$ | OFF | 1-4 |
| $DU_1(HS_1)$ | $CT_{2B}$ | OFF | ON |
| $DU_2(HS_2)$ | $CT_{1B}$ | OFF | ON |
| $DU_1(HS_1)$ | $DU_2(HS_2)$ | OFF | ON |
| $DU_2(HS_2)$ | $DU_1(HS_1)$ | OFF | ON |
| $S_1(SH_1)$ | $H1_2$ | 1-4 | ON |
| $S_2(SH_2)$ | $H1_1$ | 1-4 | ON |
| $S_1(SH_1)$ | $CT_{2A}$ | 1-4 | ON |
| $S_2(SH_2)$ | $CT_{1A}$ | 1-4 | ON |
| $CT_{1B}$ | $CT_{2B}$ | 1-4 | ON |

**Table S3.** Summary of Experimental and computed redox potential and binding free energies of tested proteins.

| PDB ID | Experiment | | Dummy-atom based TI (This study) | | Dummy-atom based CGI (Previous Study) | |
|---|---|---|---|---|---|---|
| | Redox Pot. (mV) | $\Delta G$ (kJ/mol) | Redox Pot. (mV) | $\Delta G$ (kJ/mol) | Redox Pot. Approx. (mV) | $\Delta G$, Approx. (kJ/mol) |
| 1FOV | -198 | 38.21 | -62 (3.27) | 12.01 (0.63) | -115 | 22.19 |
| 3C1R | -233 | 44.96 | -118 (6.48) | 23.04 (1.25) | -140 | 27.02 |
| 2O7K | -268 | 51.72 | -185 (1.87) | 35.62 (0.36) | -105 | 20.26 |
| 1XOA | -270 | 52.10 | -190 (5.08) | 36.58 (0.98) | -125 | 24.12 |
| 7Q53 | -302 | 58.27 | -289 (5.03) | 55.79 (0.97) | n/a | n/a |

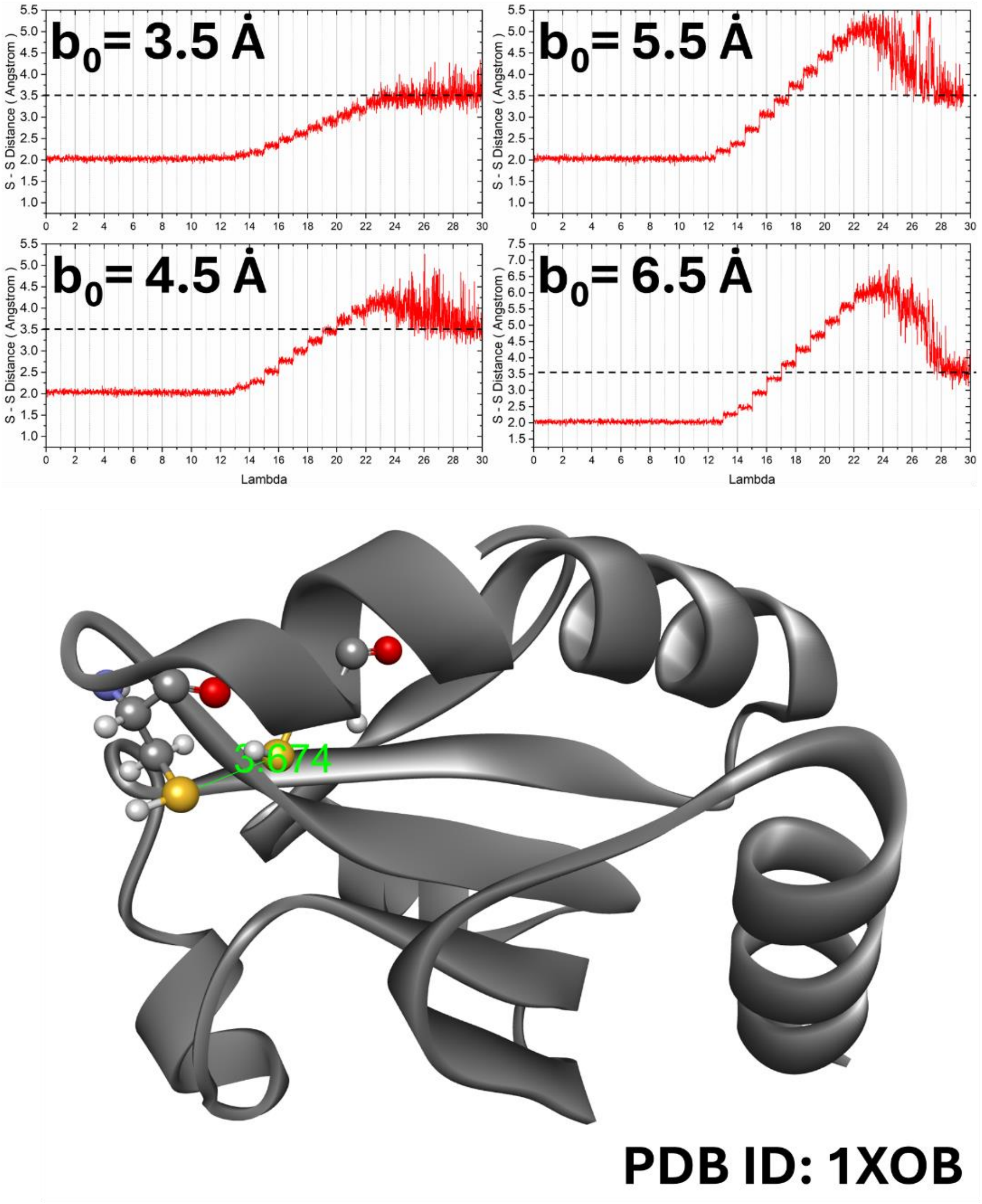

**Figure S1. (top)** Changes in distance between two sulfur atoms in 1XOA during dummy-atom based TI with different $B_0$ values for S–S distance, highlighting convergence to approx. 3.6 Å when disulfide-bonded CYSs are reduced. **(bottom)** S–S distance (3.674 Å) of the two CYSs in the crystal structure with reduced cysteines (PDB ID: 1XOB)